# Understanding of GRB-SN Connection by General Relativistic MHD Simulations

S. Nagataki

*Yukawa Institute for Theoretical Physics, Kyoto University, Oiwakecho, Kitashirakawa, Sakyoku, Kyoto 606-8502, Japan*

**Abstract.** I have developed two numerical codes to investigate the dynamics of collapsars. One is two-dimensional MHD code that are performed using the Newtonian (ZEUS-2D) code where realistic equation of state, neutrino cooling and heating processes, magnetic fields, and gravitational force from the central black hole and self-gravity are taken into account. The other one is two-dimensional general relativistic magnetohydrodynamic (GRMHD) code. At present, no microphysics is included in the GRMHD code. I have performed numerical simulations of collapsars using these codes and realistic progenitor models. In the Newtonian code, it is found that neutrino heating processes are not efficient enough to launch a jet in this study. It is also found that a jet is launched mainly by toroidal fields that are amplified by the winding-up effect. However, since the ratio of total energy relative to the rest-mass energy in the jet is not as high as several hundred, we conclude that the jets seen in this study are not GRB jets. In the GRMHD simulation, it is shown that a jet is launched from the center of the progenitor. We also find that the mass accretion rate after the launch of the jet shows rapid time variability that resembles to a typical time profile of a GRB. Even at the final stage of the simulation, bulk Lorentz factor of the jet is still low, and total energy of the jet is still as small as $10^{48}$ erg. However, we find that the energy flux per unit rest-mass flux is as high as $10^2$ at the bottom of the jet. Thus we conclude that the bulk Lorentz factor of the jet can be potentially high when it propagates outward. We also performed two-dimensional relativistic hydrodynamic simulations in the context of collapsar model to investigate the explosive nucleosynthesis happened there. It is found that the amount of $^{56}$Ni is very sensitive to the energy deposition rate. This result means that the amount of synthesized $^{56}$Ni can be little even if the total explosion energy is as large as $10^{52}$ erg. Thus, some GRBs can associate with faint supernovae. Thus we consider it is quite natural to detect no underlying supernova in some X-ray afterglows.

**Keywords:** Black Holes, General Relativity, MHD, Nucleosynthesis
**PACS:** 04.25.dg, 04.00.00, 52.30.Cv, 26.20.-f

## INTRODUCTION

Gamma-Ray Bursts (GRBs; in this study, we consider only long GRBs, so we refer to long GRBs as GRBs hereafter) have been mysterious phenomena since their discovery in 1969. Last decade, observational evidence for supernovae (SNe) and GRBs association has been reported. Some of the SNe that associate with GRBs were very energetic and blight. The estimated explosion energy was of the order $10^{52}$ ergs, and produced nickel mass was about 0.5 $M_\odot$. Thus they are categorized as a new type of SNe (sometimes called as hypernovae). The largeness of the explosion energy is very important, because it can not be explained by the standard core-collapse SN scenario, and other mechanism should be working at the center of the progenitors.

The promising scenarios are the collapsar scenario and the magnetar scenario. In the collapsar scenario, a rapidly rotating black hole (BH) is formed at the center, while a rapidly rotating neutron star with strong magnetic fields ( about $10^{15}$G) is formed in the magnetar scenario. In this study, we investigate the collapsar scenario. In the collapsar scenario, a BH is formed as a result of gravitational collapse. Also, rotation of the progenitor plays an essential role. Due to the rotation, an accretion disk is formed around the equatorial plane. On the other hand, the matter around the rotation axis falls into the BH almost freely. It is pointed out that the jet-induced explosion along the rotation axis may occur due to the heating through pair annihilation of neutrinos and anti-neutrinos that are emitted from the accretion disk. Effect of extraction of rotation energy from the accretion disk by magnetic field lines that leave the disk surface (Blandford-Payne effect) is also investigated by several authors. Recently, the effect of extraction of rotation energy from the BH through outgoing poynting flux (Blandford-Znajek effect) is investigated [1]. In order to investigate the collapsar scenario completely, a high quality numerical code including effects of a lot of microphysics (neutrino physics, nuclear physics, and equation of state for dense matter) and macrophysics (magneto-hydrodynamics, general relativity) has to be developed. Although many numerical studies have been reported, such a numerical code has not been developed yet. Thus we have to develop our numerical code step by step.

In this paper, we investigate the dynamics of collapsars using two numerical codes that I have developed. One is two-dimensional MHD code that are performed using the Newtonian (ZEUS-2D) code where realistic equation of state, neutrino cooling and heating processes, magnetic fields, and gravitational force from the central black hole and self-gravity are taken into account [2]. The other one is two-dimensional general relativistic magnetohydrodynamic (GRMHD) code [3]. At present, no microphysics is included in the GRMHD code. I have performed numerical simulations of collapsars using these codes and realistic progenitor models.

As a result, I found the following points: In the Newtonian code, it is found that neutrino heating processes are not efficient enough to launch a jet in this study. It is also found that a jet is launched mainly by toroidal fields that are amplified by the winding-up effect. However, since the ratio of total energy relative to the rest-mass energy in the jet is not as high as several hundred, we conclude that the jets seen in this study are not GRB jets. In the GRMHD simulation, it is shown that a jet is launched from the center of the progenitor. We also find that the mass accretion rate after the launch of the jet shows rapid time variability that resembles to a typical time profile of a GRB. Even at the final stage of the simulation, bulk Lorentz factor of the jet is still low, and total energy of the jet is still as small as $10^{48}$ erg. However, we find that the energy flux per unit rest-mass flux is as high as $10^2$ at the bottom of the jet. Thus we conclude that the bulk Lorentz factor of the jet can be potentially high when it propagates outward.

Also, we perform two-dimensional relativistic hydrodynamic simulations in the context of collapsar model to investigate the explosive nucleosynthesis happened there. We have to be careful to the point that where and when $^{56}$Ni is synthesized in the collapsar is not unclear. This is because the explosion mechanism is not known well. One possibility is that $^{56}$Ni is synthesized in the jet region like jet-like supernova explosion [4, 5]. Another possibility is that nucleosynthesis in the accretion disk, and some fraction of the accreting matter escape from the system [6]. In this paper, we consider the former possibility.

As a result, it is found that the amount of $^{56}$Ni is very sensitive to the energy deposition rate. This result means that the amount of synthesized $^{56}$Ni is little even if the total explosion energy is as large as $10^{52}$ erg. Thus, some GRBs can associate with faint supernovae. Thus we consider it is quite natural to detect no underlying supernova in some X-ray afterglows.

# NEWTONIAN SIMULATIONS

## Formulation

We perform two-dimensional MHD simulations taking account of self-gravity and gravitational potential of the central point mass. The calculated region corresponds to a quarter of the meridian plane under the assumption of axisymmetry and equatorial symmetry. The radial grid is nonuniform, extending from $10^6$ to $10^{10}$ cm with finer grids near the center, while the polar grid is uniform. We adopt the model E25 in Heger et al. (2000). This model corresponds to a star that has 25 $M_\odot$ initially with solar metallicity but loses its mass and becomes 5.45 $M_\odot$ of a Wolf-Rayet star at the final stage.

The EOS used in this study is the one developed by [7]. This EOS contains an electron-positron gas with arbitrary degeneracy, which is in thermal equilibrium with blackbody radiation and ideal gas of nuclei. Although the ideal gas contribution of nuclei to the total pressure is negligible, effects of energy gain/loss due to nuclear reactions are important. In this study, nuclear statistical equilibrium is assumed for the region where $T > 5 \times 10^9$ K is satisfied.

Neutrino cooling processes due to pair capture on free nucleons, pair annihilation, and plasmon decay are included in this study. The neutrino heating process due to neutrino pair annihilation and electron-type neutrino captures on free nucleons with blocking factors of electrons and positrons are included in this study.

## Results

### Dynamics without Magnetic Fields

We show in Figure 1 the density contour of the central region of the progenitor ($r < 10^8$ cm) with velocity fields at 2.2 sec for the case without magnetic fields. The contour represents the density (g/cc) in logarithmic scale ($10^3$-$10^{12}$). Vertical axis and horizontal axis represent polar axis (=rotation axis) and equatorial plane, respectively. In the case without magnetic fields, no jet is seen even if neutrino pair-annihilation effects are taken into account in this simulation.

### Dynamics with Magnetic Fields

When vertical, weak magnetic fields are put as an initial condition, the dynamics is changed. An example is shown in Figure 2 where initial amplitude of the magnetic fields is set to be $10^9$G. It is found that a jet is launched by magnetic fields (in particular, toroidal mag-netic fields that are amplified by the winding-up effect). However, the ratio of total energy relative to the rest-mass energy in the jet at the final stage of simulations suggests that the bulk Lorentz factor of the jet will not reach as high as several hundred, so it is concluded that the jet seen in this study will not be a GRB jet.

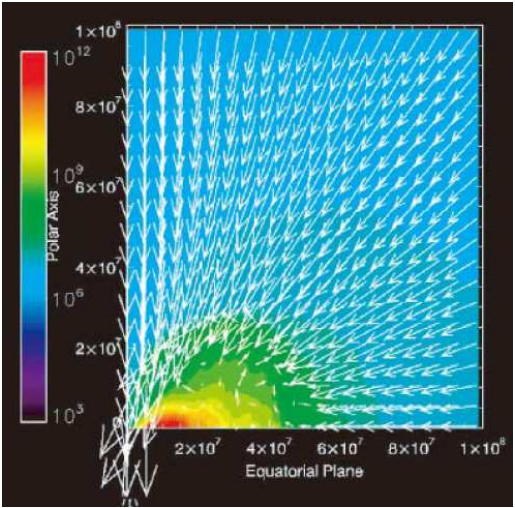

**FIGURE 1.** Density contour of the central region of the progenitor ($r < 10^8$ cm) with velocity fields at 2.2 sec for the case without magnetic fields. The contour represents the density (g/cc) in logarithmic scale ($10^3$-$10^{12}$). Vertical axis and horizontal axis represent polar axis (=rotation axis) and equatorial plane.

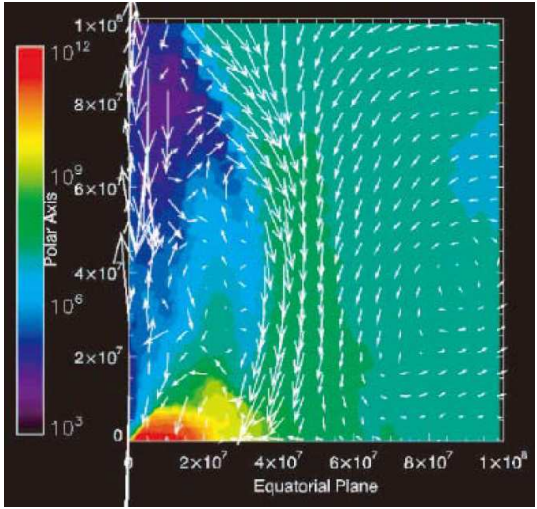

**FIGURE 2.** Same with Figure 1, but with magnetic fields. Initial amplitude of the vertical magnetic fields is set to be $10^9$G.

# GENERAL RELATIVISTIC MHD SIMULATIONS

## Formulation

We have developed a two-dimensional GRMHD code following [8, 9]. We have adopted a conservative, shock-capturing scheme with Harten, Lax, and van Leer (HLL) flux term with flux-interpolated constrained transport technique. We use a third-order Total Variation Diminishing (TVD) Runge-Kutta method for evolution in time, while monotonized central slope-limited linear interpolation method is used for second-order accuracy in space. 2D scheme (2-dimensional Newton-Raphson method) is usually adopted for transforming conserved variables to primitive variables.

When we perform simulations of GRMHD, Modified Kerr-Schild coordinate is basically adopted with mass of the BH (M) fixed where the Kerr-Schild radius r is replaced by the logarithmic radial coordinate $x1 = \ln r$. When we show the result, the coordinates are sometimes transfered from Modified Kerr-Schild coordinate to Kerr-Schild one for convenience. In the following, we use G $=M = c = 1$ unit. G is the gravitational constant, c is the speed of light, and M is the gravitational mass of the BH at the center.

The calculated region covers from $r = 1.8$ to $3 \times 10^4$ (that corresponds to $5.3 \times 10^5$ cm and $8.9 \times 10^9$ cm in cgs units) with uniform grids in the Modified Kerr-Schild space.

We adopt the model 12TJ in [10]. This model corresponds to a star that has 12 $M_\odot$ initially with 1% of solar metallicity, and rotates rapidly and does not lose its angular momentum so much by adopting small mass loss rate. As a result, this star has a relatively large iron core of 1.82 $M_\odot$, and rotates rapidly (the estimated Kerr parameter that a BH forming of mass and angular momentum of the inner 3 $M_\odot$ would formally have is 0.57) at the final stage. Thus we set the Kerr-parameter of the black hole at the center is set to be 0.5 throughout of the simulation.

The initial, weak poloidal magnetic fields are put initially. The minimum plasma beta in the simulation region is greater than 100 initially.

## Results

In Figure 3, color contours of rest mass density at the central region are shown. Colors represent the density in units of g cm$^{-3}$ in logarithmic scale. The length r = 200 corresponds to 5.9 times $10^7$ cm. The time unit corresponds to 9.85 times $10^{-6}$ sec. Figure 3. shows the contours at t = 180000 (that corresponds to 1.773 sec). A jet is clearly seen along the rotation axis.

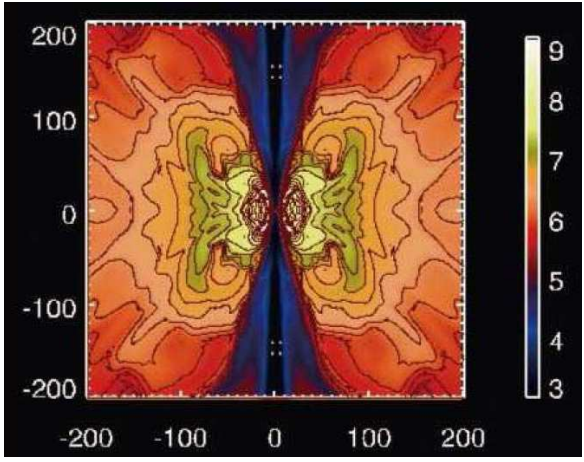

**FIGURE 3.** Contours of rest mass density at the central region in logarithmic scale at t = 180000 (that corresponds to 1.773 sec), in which cgs units are used assuming that the gravitational mass of the BH is 2 $M_\odot$. The length unit in the vertical/horizontal axes corresponds to 2.95 times$10^5$ cm.

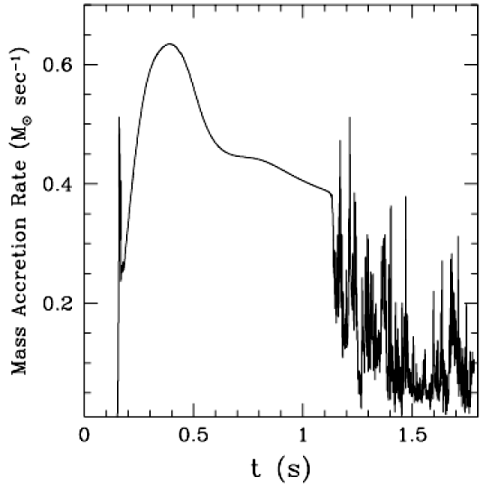

**FIGURE 4.** Mass accretion rate history on the horizon.

In Figure 4, mass accretion rate history on the horizon is shown. It takes about 0.15 sec for the inner edge of the matter to reach the horizon. When the matter reaches there, there is an initial spike of the mass accretion rate. After that, there is a quasi-steady state is realized. Then, the jet is launched at about 1.1 sec. After that, the mass accretion rate varies rapidly with time, which resembles to a typical time profile of a GRB.

Figure 5 shows color contours of the plasma beta (pgas/pmag) in logarithmic scale at t = 180000. As expected, the plasma beta is low in the jet region while it is high in the accretion disk region. Color contours of the energy flux per unit rest mass flux, which is conserved for an inviscid fluid flow of magnetized plasma, are also shown in Figure 6 (in logarithmic scale). This value represents the bulk Lorentz factor of the invischid fluid element when all of the internal and magnetic energy are converted into kinetic energy at large distances. We can see that the bulk Lorentz factor of the jet can be potentially as high as $10^2$ at large radius.

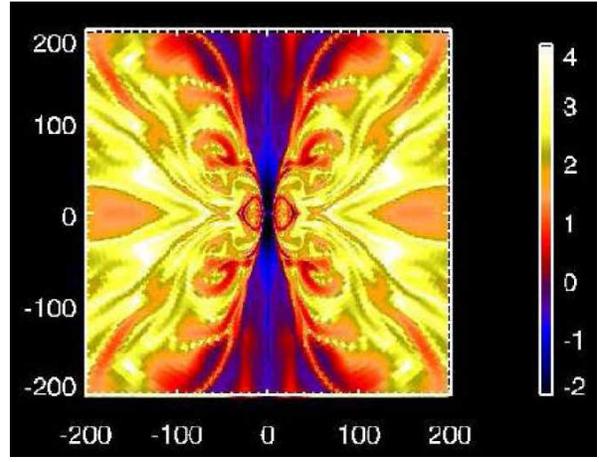

**FIGURE 5.** Contour of the plasma beta (pgas/pmag) at t = 180000 in logarithmic scale.

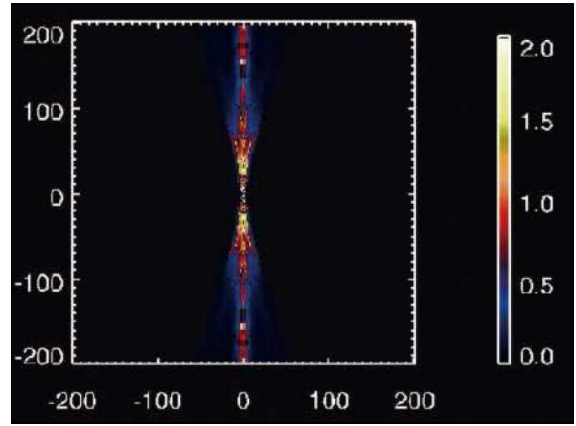

**FIGURE 6.** Contours of the energy flux per unit rest-mass flux at t = 180000 that represent the bulk Lorentz factor of the invischid fluid element when all of the internal and magnetic energy are converted into kinetic energy at large distances. The contours are written in logarithmic scale.

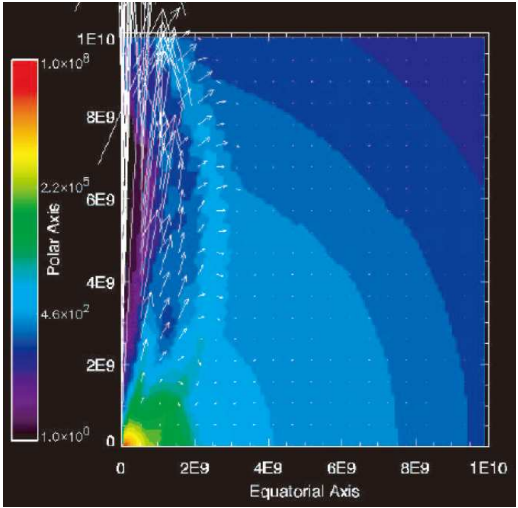

**FIGURE 7.** Density structure and velocity field of model E51 at 1.5 sec.

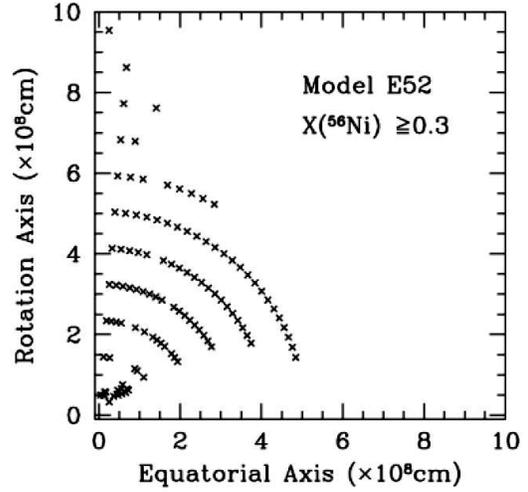

**FIGURE 9.** Same as Figure 8, but for model E52. The total ejected mass of $^{56}$Ni is 0.23 solar masses.

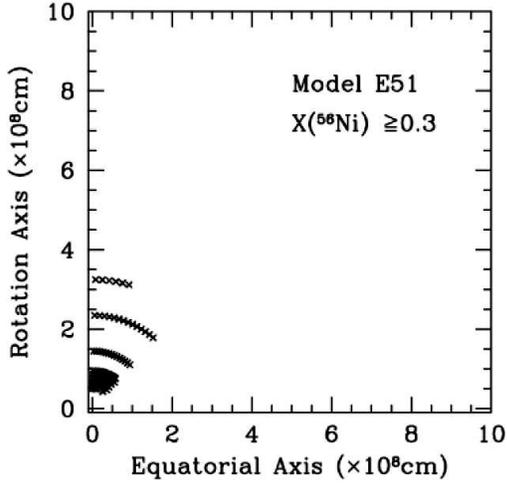

**FIGURE 8.** Positions of the ejected test particles at 0 sec that meet the condition that the mass fraction of $^{56}$Ni becomes greater than 0.3 as a result of explosive nucleosynthesis for model E51. The total ejected mass of $^{56}$Ni is 0.0439 $M_\odot$.

## EXPLOSIVE NUCLEOSYNTHESIS IN COLLAPSAR

### Formulation

We have done two-dimensional hydrodynamic simulations taking into account self-gravity and the gravitational potential of the central point mass. The calculated region corresponds to a quarter of the meridian plane under the assumption of axisymmetry and equatorial symmetry. The spherical mesh with 250 by 30 grid points is used for all the computations. The radial grid is nonuniform, extending from $2 \times 10^7$ to $3 \times 10^{11}$ cm with finer grids near the center, while the polar grid is uniform.

We adopt the collapsar model of MacFadyen & Woosley (1999). When the central black hole has acquired a mass of 3.762 $M_\odot$, we map the model to our computational grid. The surface of the helium star is 2.98 times $10^{10}$ cm. Electron fraction, Ye, is set to be 0.5 throughout of this paper since neutrino process is not included.

To simulate the jet-induced explosion, we deposit only thermal energy at a rate $10^{51}$ ergs/s homogeneously within a 30 degree cone around the rotation axis for 10 sec. In the radial direction, the deposition region extends from the inner grid boundary located at 200 km to a radius of 600 km. This treatment is same as that of [11]. We name this model E51. We consider this model to be the standard one. For comparison, we perform a calculation in which total explosion energy ($10^{52}$ ergs) is put initially with the same deposition region as model E51. We name these models E52. We consider that this model represents an extreme case.

Since the hydrodynamics code is Eulerian, we use the test particle method [12, 13] in order to obtain the information on the time evolution of the physical quantities along the fluid motion, which is then used for the calculations of the explosive nucleosynthesis. Test particles are scattered in the progenitor and are set at rest initially. They move with the local fluid velocity at their own positions after the passage of the shock wave. The temperature and density that each test particle experiences at each time step are preserved.

Since the chemical composition behind the shock wave is not in nuclear statistical equilibrium, the explosive nucleosynthesis has to be calculated using the time evolution of density and temperature, and a nuclear reaction network, which is called post-processing. The nuclear reaction network contains 250 species.

## Results

We deposited thermal energy to launch a jet from the central region of the collapsar. The density structure for models E51 at 1.0 sec is shown in Figure 7. It is clearly shown that a sharp, narrow jet propagates along the rotation axis in model E51, which is similar to [11]. On the other hand, in the case of E52, a broad, deformed shock wave propagates in the progenitor (see also [5]).

In Figure 8, positions of the ejected test particles for model E51 at 0 sec are shown that satisfy the condition that the mass fraction of $^{56}$Ni becomes greater than 0.3 as a result of explosive nucleosynthesis. The total ejected mass of $^{56}$Ni becomes 0.0439 $M_\odot$, which is much smaller than the observed values of hypernovae. In Figure 9, the same values are shown as in Figure 8, but for models E52. The total ejected mass of $^{56}$Ni is 0.23 $M_\odot$, which is comparable to the observed values of hypernovae.

Thus we can conclude that the resulting amount of $^{56}$Ni is very sensitive to the energy deposition rate. This result means that the amount of synthesized $^{56}$Ni can be little even if the total explosion energy is as large as $10^{52}$ erg. Thus, some GRBs can associate with faint supernovae. Thus we consider it is quite natural to detect no underlying supernova in some X-ray afterglows such as GRB060614.

## SUMMARY AND CONCLUSION

I have developed two numerical codes to investigate the dynamics of collapsars. One is two-dimensional MHD code that are performed using the Newtonian (ZEUS-2D) code where realistic equation of state, neutrino cooling and heating processes, magnetic fields, and gravitational force from the central black hole and self-gravity are taken into account. The other one is two-dimensional general relativistic magnetohydrodynamic (GRMHD) code. I have performed numerical simulations of collapsars using these codes and realistic progenitor models. In the Newtonian code, it is found that neutrino heating processes are not efficient enough to launch a jet in this study. It is also found that a jet is launched mainly by toroidal fields that are amplified by the winding-up effect. However, since the ratio of total energy relative to the rest-mass energy in the jet is not as high as several hundred, we conclude that the jets seen in this study are not GRB jets. In the GRMHD simulation, it is shown that a jet is launched from the center of the progenitor. We also find that the mass accretion rate after the launch of the jet shows rapid time variability that resembles to a typical time profile of a GRB. Even at the final stage of the simulation, bulk Lorentz factor of the jet is still low, and total energy of the jet is still as small as $10^{48}$ erg. However, we find that the energy flux per unit rest-mass flux is as high as $10^2$ at the bottom of the jet. Thus we conclude that the bulk Lorentz factor of the jet can be potentially high when it propagates outward. We also performed two-dimensional relativistic hydrodynamic simulations in the context of collapsar model to investigate the explosive nucleosynthesis happened there. It is found that the amount of $^{56}$Ni is very sensitive to the energy deposition rate. This result means that the amount of synthesized $^{56}$Ni can be little even if the total explosion energy is as large as $10^{52}$ erg. Thus, some GRBs can associate with faint supernovae. Thus we consider it is quite natural to detect no underlying supernova in some X-ray afterglows such as GRB060614.